\documentstyle[prl,aps,preprint]{revtex}
\begin{document}
\draft
\title{Insecurity of Quantum Secure Computations}
\author{Hoi-Kwong Lo\footnotemark}
\address{BRIMS, Hewlett-Packard Labs, \\
Filton Road, Stoke Gifford \\
Bristol, BS12 6QZ\\
U.K.
}
\author{and}
\address{
Institute for Theoretical Physics \\
University of California, Santa Barbara \\
CA 93106-4030 \\
U. S. A.
}

\date{\today}
\footnotetext[1]{hkl@hplb.hpl.hp.com}
\maketitle
\begin{abstract}
It had been widely claimed that quantum mechanics can
protect private information during public decision
in for example the so-called two-party secure computation.
If this were the case, quantum
smart-cards could prevent fake teller machines
from learning the PIN
(Personal Identification Number) from the customers' input.
Although such optimism has been challenged by
the recent surprising discovery of the insecurity of
the so-called quantum bit commitment,
the security of quantum two-party computation
itself remains unaddressed.
Here I answer this question directly by
showing that all {\it one-sided} two-party computations
(which allow only one of
the two parties to learn the result)
are necessarily insecure.
As corollaries to my results, quantum one-way oblivious
password identification and the so-called quantum one-out-of-two oblivious
transfer are impossible.
I also construct a class of functions
that cannot be computed
securely in any {\it two-sided} two-party computation.
Nevertheless, quantum cryptography remains
useful in key distribution and can still provide
partial security in ``quantum money'' proposed
by Wiesner.

\end{abstract}
\narrowtext
\section{Introduction}
\label{Intro}
Copying
of an unknown quantum state (by for example an
eavesdropper) is strictly forbidden
by the linearity of quantum mechanics\cite{WZ}.
Consequently, quantum
cryptography\footnote{Quantum Cryptography was first proposed
by Wiesner\cite{Wiesn} in about 1970 in a manuscript that
remained unpublished until 1983.} (or more
precisely quantum key
distribution\cite{Collins,BBE,BB84,Ekert,B92})
allows two users
to share a common random secret string of information which can then
be used to make their subsequent communications totally unintelligible to
an eavesdropper. In this paper I am, however, concerned with another class
of applications of quantum cryptography---the protection of
private information during public decision\cite{Brass,Brass2}.
For instance, two millionaires may be interested in
knowing who is richer but neither wishes to disclose the precise
amount of money that he/she has.
More generally, in a {\it one-sided} two-party computation,
Alice has a private
input $i$ and Bob a private input $j$.
Alice would like to help Bob to compute a prescribed
function $f(i,j)$ without revealing anything about $i$
more than what is logically necessary. (For a precise
definition of a one-sided two-party computation, see Section~2.)
In classical cryptography, such two-party computations
can be made secure only either 1) through
trusted intermediaries or 2) by accepting some
unproven computational
assumptions.\footnote{In the first case, if both Alice and
Bob trust Charles, they
simply tell him their private inputs and let Charles
perform the computation on their behalf and tell them the result
afterwards. The problem here
is that Charles can cheat by telling either Alice or Bob the other
party's private input. In the second case, assumptions such as
the hardness of factoring can be used. However,
an adversary with unlimited computing power (or with a
quantum computer\cite{Shor}) can defeat such
unproven computational assumptions.}
The impossibility of {\it unconditionally} secure two-party
computation in {\it  classical} cryptography
had
led to much interest in {\it quantum} cryptographic
protocols\cite{Wiesn,BB84,BC91,Ard,Ar,CS,Huttner,BCJL,BBCS,Yao}
which are supposed to be unconditionally
secure\cite{BCJL,BBCS,Yao}.

An important primitive in secure computation is
the so-called bit commitment.\footnote{The basic idea of
bit commitment is to conceal information and to reveal
it later. It might be useful
to note that
Yao\cite{Yao} has shown that any secure quantum bit commitment scheme
can be used to implement secure quantum oblivious transfer
whereas Kilian\cite{Kilian} has shown that, in classical cryptography,
oblivious transfer can be used to implement two-party secure
computation. Therefore, this chain
of argument appears to suggest that, with quantum bit
commitment, quantum
cryptography could achieve unconditionally
secure two-party computation, thus solving a long
standing problem in cryptography.}
The optimism in unconditional secure quantum two-party computation
was largely contributed by
well-known claims of unconditional secure
quantum bit commitment protocols\cite{BCJL} (and also
oblivious transfer\cite{BBCS,Yao}).
However, such optimism has recently been put into serious question 
due to the surprising demonstration of the insecurity of
quantum
bit commitment (against an EPR-type of
attack with delayed measurements) by
Mayers\cite{Mayers1,Mayers2} and also by
Chau and me\cite{LC1,LC2}.
Yet an important
question remains: Other than quantum key distribution, can
quantum cryptographic protocols, in particular, two-party
computation, be unconditionally
secure at all? This is an important question because,
in many cases,
quantum bit commitment might be thought of as a means to
an end---two party secure computation. If
secure quantum two-party computation is possible,
many applications of quantum cryptography,
such as the prevention of frauds due to typing PIN
(Personal Identification Number) to
dishonest teller machine mentioned in the abstract,
will still
survive.

Amazingly, one possible viewpoint to take is that there is really nothing to
prove because the standard reduction
theorems\cite{Kilian,Crepe,CreKi}\footnote{I thank G. Brassard for
helpful discussions about those standard reduction theorems.}
in classical cryptography immediately imply
that quantum one-sided two-party computation
is impossible: In classical cryptography, an example of
one-sided two party computation is one-out-of-two
oblivious transfer, which can be used to implement bit commitment.
If bit commitment is impossible, one-sided two-party computations
must also generally be impossible.
Doubt has been expressed in the literature concerning
the validity of this standard
reduction in a quantum model\cite{Brass2}.
Here I argue that by definition the standard
reduction must apply
to quantum cryptographic protocols: Bit commitment,
oblivious transfer and two-party computations are classical concepts
whose
security requirements are defined in a classical probabilistic language.
If there is any sense at all in saying that a quantum protocol can
achieve say two-party computation, it is {\it a matter of definition}
that the quantum protocol has to
satisfy the classical probabilistic security requirements under
all circumstances.
In particular, one must be allowed to use a quantum
cryptographic protocol as a ``black box'' primitive
in building up more sophisticated protocols and to analyze the
security of those new protocols with {\it classical} probability
theory.\footnote{One may get the
feeling from reading the literature that a
quantum protocol should be regarded
as secure if it appears to satisfy
its security requirements when it is executed {\it only once} and {\it in
isolation}.  This, however, does {\it not} guarantee that it
satisfies the
security requirements when it is used as a subroutine
of a larger routine because a cheater might defeat the security of
the larger routine by performing coherent measurements.
Therefore, I think that a more accurate definition of a secure
quantum protocol should be much more stringent.}

By adopting this new and, in my opinion,
more accurate definition of secure quantum protocols,
one sees that the impossibility of quantum bit commitment
immediately implies the impossibility of quantum one-sided two-party
computations (and one-out-of-two oblivious transfer
as well as oblivious transfer) and this is the end of the story.

Yet such an ending is disappointing in two aspects.
While such a viewpoint is conceptually correct, it is a bit
formal and non-constructive. A constructive proof
would make things more transparent and convincing.
A perhaps more serious objection is that
while such an argument rules out one-out-of-two
oblivious transfer and the two-party computation of
a general function, there remains the
possibility that {\it some} special class of
functions (whose two-party computations
cannot be
used to implement one-out-of-two
oblivious transfer\footnote{According to Kilian, such functions
do exist.}) might still be computed securely in
one-sided two-party computations.
Here I
investigate directly the security of one-sided
two-party computation without
using the formal standard reduction. My main result is
that one-sided quantum two party
secure computation is always impossible.\footnote{Remarkably,
an alternative proof of the impossibility of {\it ideal}
quantum one-sided two party computation
can be made by generalizing
Wiesner's\cite{Wiesn} early insight on the impossibility of one-way
scheme for so-called one-out-of-two oblivious transfer
and combining it with the idea of the proof of
the impossibility of quantum bit commitment. I omit this alternative proof
here because it is not transparent at all.}
(For its definition, see Section~2.)
That is to say that, as far as one-sided two-party computations
are concerned, quantum cryptography
is absolutely useless. As a corollary, the so-called
quantum one-out-of-two oblivious
transfer is also impossible. I also present
a class of functions that {\it cannot} be computed in any {\it two-sided}
two-party computation.
Nevertheless, quantum cryptography remains useful for key distribution
and can still provide partial security in ``quantum money'' proposed by
Wiesner.

\section{ideal one-sided two-party secure computation}
\label{main}
\subsection{Definition and Security Requirements}
Suppose Alice has a private (i.e., secret) input
$i \in \{1, 2, \cdots , n \}$ and
Bob has a private input $j \in \{1, 2, \cdots , m \}$.
An {\it ideal} one-sided two party secure computation is defined as follows:
Alice helps Bob to
compute a prescribed function
$f(i,j) \in \{1,2, \cdots , p \}$ in
such a way that at the end of the protocol,

(a) Bob learns $f(i,j)$ unambiguously,

(b) Alice learns nothing (about $j$ or $f(i,j)$), 

and

(c) Bob knows nothing about $i$ more than
what logically follows from the values of $j$ and $f(i,j)$.

Notice that, for a one-sided two-party computation protocol to
be secure, Bob is supposed to input a {\it particular}
value of $j$ and to learn the value of $f(i,j)$ for
that particular value of $j$ {\it only}.
I will show that these three security requirements (a), (b) and (c)
are incompatible
in the following manner: Assuming that the first two security
requirements (a) and (b) are satisfied, I will work out a cheating
strategy for Bob which would allow him to learn
the values of $f(i,j)$ for {\it all} $j$'s, thus
violating security requirement (c).\footnote{In other words,
instead of the ideal one-sided two-party secure computation protocol,
quantum cryptography gives only a protocol that
allows Bob to learn $f(i,j)$ for {\it all}
$j$'s. Such a protocol is uninteresting as it can be
achieved in classical cryptography simply by having Alice
tell Bob those values. Therefore, quantum
cryptography provides no real advantage in this ideal case.}
I, therefore, conclude that
ideal quantum one-sided two-party computations are impossible.
In Section~4, I will generalize this result to non-ideal
protocols (which may violate security requirements (a) and (b) slightly).

\subsection{Bob's cheating strategy}
Consider the following cheating strategy by Bob
who determines the values of $f(i,j_1),
f(i,j_2), \cdots , f(i,j_m)$ successively:
Bob first inputs a
value $j_1$ for $j$ and goes through the protocol.
At the end of the protocol, he determines the value of
$f(i,j_1)$.
He then applies a unitary transformation to change the value of $j$ from
$j_1$ to $j_2$ and determines $f(i,j_2)$.
After that,
Bob applies a unitary transformation to change $j$ from
$j_2$ to $j_3$ and determines $f(i,j_3)$
and so on.

\subsection{Key Points of the Proof}
The above cheating strategy by Bob works for two
reasons. First, using the insight gained from the
impossibility of quantum bit commitment\cite{Mayers1,Mayers2,LC1,LC2},
in Subsection~3B I will
prove the following: The
security requirement (b)---that Alice knows nothing about
$j$---implies that at the end of the protocol, Bob can cheat by changing the
value of $j$ from $j_1$ to $j_2$ by applying a unitary transformation
to his own quantum machine.\footnote{The impossibility of
quantum bit commitment\cite{Mayers1,Mayers2,LC1,LC2}
essentially states that if
Alice does not know something, then Bob can change it.
The commitment made by Bob is, therefore, fake.} Consequently, Bob can
determine the value of $f(i,j_2)$ instead of $f(i,j_1)$, as long
as he has {\it not} measured $f(i,j_1)$ yet. Of course, Bob would
like to learn $f(i,j_1)$ and he {\it does} measure $f(i,j_1)$ before
rotating $j_1$ to $j_2$. At first sight, this seems to be
a problem because measurements in quantum mechanics generally
disturb a signal.
Here comes the second point.
Measurement of $f(i,j_1)$ does not
disturb Bob's state at all for the following
reason. Since, by the security requirement (a)
of an ideal protocol, Bob can
input $j=j_1$ and learn the value of $f(i,j_1)$ unambiguously, the
density matrix that Bob has must be an eigenstate of the measurement
operator that he uses for determining $f(i,j_1)$.
Being an eigenstate, the density matrix is, therefore, undisturbed by
Bob's measurement. QED

In effect, I am arguing that the density matrix Bob has is a
simultaneous eigenstate of the measurement operators
$f(i,j_1), f(i,j_2), \cdots , f(i,j_m)$. See Subsection~3B.

\section{details of the proof}

\subsection{Unitary description}
Let me present my result in more detail.
It is convenient to use a unitary description of
two-party computation\cite{Mayers2,LC2}. Let $H_A$ ($H_B$ respectively)
denote the Hilbert space
of Alice's (Bob's) quantum machine.
Imagine a two-party computation
in which both Alice and Bob possess quantum computers and
quantum storage devices. By maintaining the quantum coherence
of the composite quantum system, $H_A \otimes H_B$, (using external
control such as classical computers,
assembling of quantum gate arrays, quantum error correction
and fault-tolerant quantum computation) one can
avoid dealing with the collapse of the wavefunction.
Alice and Bob's actions on their quantum machines can be
summarized\footnote{For the
basic idea, see \cite{Mayers2}. For detailed justification with a
concrete model (a variant of Yao's model\cite{Yao})
see \cite{LC2}.
Of course, in reality the execution of the protocol
may not require quantum computers. This is, however, equivalent to
a situation when the parties do not make full use of their quantum computers.
If one can show that a cheater can cheat successfully against
an honest party who has a quantum computer, clearly the cheater can
cheat successfully against one without. Therefore,
a unitary description is very useful for
my purposes.} as an overall unitary transformation $U$ applied
to the initial state $| u \rangle_{in} \in H_A \otimes H_B$.
i.e., 
\begin{equation}
| u \rangle_{fin} = U | u \rangle_{in}.
\label{inout}
\end{equation}
The unitary transformation, $U$, is known to both Alice and Bob
because they know the procedure of the protocol.
When both
parties are honest, $| u^h \rangle_{in} =| i \rangle_A \otimes | j \rangle_B $ 
and
\begin{equation}
| u^h\rangle_{fin} = | v_{ij} \rangle
\equiv U \left( | i \rangle_A \otimes | j \rangle_B
\right).
\label{vij}
\end{equation}
Therefore, the density matrix that Bob has at the end of
protocol is simply
\begin{equation}
\rho^{i,j} = {\rm Tr}_A | v_{ij} \rangle
\langle  v_{ij} | .
\label{rhoij}
\end{equation}

\subsection{Changing $j$ from $j_1$ to $j_2$}

I asserted in the last section that, owing to the security
requirement (b), at the end of the protocol Bob can change the value
of $j$ from $j_1$ to $j_2$ by applying a unitary transformation
to the state of his quantum machine. Since the value of Alice's input $i$
is unknown to Bob, for such a cheating strategy to work,
I need to prove that
this unitary transformation can be chosen
to be independent of the value of
$i$\footnote{Using the idea of the impossibility of bit
commitment\cite{Mayers1,Mayers2,LC1,LC2},
it is trivial to prove that,
for {\it each} $i$, a unitary
transformation $U^{i, j_1, j_2}$ that rotates $j$ from $j_1$ to
$j_2$ exists.
What is less trivial to prove is the
existence of a unitary
transformation $U^{j_1, j_2}$ which works for {\it all} $i$'s
simultaneously. I thank D. Mayers for enlightening
discussions.

Actually, Bob can choose his unitary transformation
according to the outcome of his measurement. This observation
will be useful in later discussion
just before Corollary A in the next Section.}:

{\it Assertion:} Given $j_1, j_2 \in \{1, 2, \cdots , m\}$,
there exists a unitary transformation $U^{j_1, j_2}$ such that
\begin{equation}
U^{j_1, j_2} \rho^{i,j_1 } \left( U^{j_1, j_2} \right)^{-1} =
\rho^{i,j_2 }
\label{rot}
\end{equation}
for {\it all} $i$.

{\it Proof}: Notice that Bob must allow
Alice to choose the value of her input, $i$, randomly.
But then a dishonest Alice may try to learn about $j$
by an EPR-type of attack. i.e., she
entangles the state of her quantum machine $A$ with her
quantum dice $D$ and prepares the initial state
\begin{equation}
{ 1 \over \sqrt{n}} \sum_i | i \rangle_D \otimes | i \rangle_A .
\label{Alicec}
\end{equation}
(Recall that $n$ is the cardinality of $i$.)
Instead of measuring the state of her quantum dice $D$ honestly,
she may keep $D$ for herself and
use the second register, $A$, to execute the two party protocol
honestly from this point on.
Suppose Bob's input is $j_1$.
The initial state is, therefore,
\begin{equation}
|u' \rangle_{in} = { 1 \over \sqrt{n}}
\sum_i | i \rangle_D \otimes | i \rangle_A
\otimes |j_1 \rangle_B .
\label{initialc}
\end{equation}
At the end of the protocol, it follows from
Eqs.\ (\ref{inout}) and (\ref{initialc}) that
the total
wave function of the combined system $D$, $A$ and $B$
is described by
\begin{equation}
|{v_j}_1 \rangle =
 {1 \over \sqrt{n}}
\sum_i | i \rangle_D \otimes  U \left( | i \rangle_A \otimes |j_1 \rangle_B
\right).
\label{vj1}
\end{equation}
Similarly, if Bob's input is $j_2$ instead,
the total wavefunction at the end of the protocol will be
\begin{equation}
|{v_j}_2 \rangle =
 {1 \over \sqrt{n}}
\sum_i | i \rangle_D \otimes  U \left( | i \rangle_A \otimes |j_2 \rangle_B
\right).
\label{vj2}
\end{equation}
An ideal protocol should prevent such a dishonest Alice
from learning anything about $j$. Therefore, the reduced density
matrices in Alice's hand for the two cases
$j=j_1$ and $j=j_2$ must be the same, i.e.,
\begin{equation}
\rho^{Alice}_{j_1} = {\rm Tr}_B |{v_j}_1 \rangle \langle {v_j}_1 | =
{\rm Tr}_B |{v_j}_2 \rangle \langle {v_j}_2 |= \rho^{Alice}_{j_2}.
\label{Alice}
\end{equation}
Equivalently, the two wavefunctions, $|{v_j}_1 \rangle$ and
$|{v_j}_2 \rangle$ have the same Schmidt decomposition\cite{Sch}.
i.e.,
\begin{equation}
|{v_j}_1 \rangle =
\sum_k  a_k | \alpha_k \rangle_{AD} \otimes | \beta_k  \rangle_B
\label{1vj1}
\end{equation}
and
\begin{equation}
|{v_j}_2 \rangle =
\sum_k  a_k | \alpha_k \rangle_{AD} \otimes | \beta'_k  \rangle_B .
\label{1vj2}
\end{equation}
Here $ | \alpha_k \rangle_{AD}$, $| \beta_k  \rangle_B$
and $| \beta'_k  \rangle_B$ are eigenvectors of the
corresponding density matrices and satisfy
$\langle \alpha_{k'}| \alpha_k \rangle_{AD} = \delta_{k,k'}$, etc.
Notice that
Eqs.\ (\ref{1vj1}) and (\ref{1vj2}) contain
the same factors $a_k$ and $| \alpha_k \rangle_{AD}$ and the only difference
lies in the state of Bob's quantum machine, $B$.
Now, consider the unitary transformation
$U^{j_1, j_2}$ that rotates
$| \beta_k  \rangle_B $ to $ | \beta'_k  \rangle_B$. Notice that it
acts on
$H_B$ {\it alone} and yet,
as can be seen from Eqs.\ (\ref{1vj1}) and (\ref{1vj2}), it
rotates $|{v_j}_1 \rangle$ to $|{v_j}_2 \rangle$.
i.e.,
\begin{equation}
|{v_j}_2 \rangle =
U^{j_1, j_2} |{v_j}_1 \rangle .
\label{roj}
\end{equation}
Since 
\begin{equation}
{}_D \langle i | v_j \rangle = { 1 \over \sqrt{n}} | v_{i j} \rangle
\label{new}
\end{equation}
(see Eqs. (\ref{vij}), (\ref{vj1}) and (\ref{vj2})), by
multiplying Eq.\ (\ref{roj}) by $ {}_D \langle i |$ on the left,
one finds that
\begin{equation}
| v_{i{j_2}} \rangle =
U^{j_1, j_2} | v_{i{j_1}} \rangle .
\label{roj1}
\end{equation}
As one is interested in Bob's reduced density matrix,
one takes the trace of
$| v_{i{j_2}} \rangle \langle  v_{i{j_2}}|$ over $H_A$
and uses Eq.\ (\ref{roj1}) to obtain Eq.\ (\ref{rot}).
This completes the proof of my assertion, Eq. (\ref{rot}).

The implication of Eq.~(\ref{rot}) is profound.
Independent of the value of Alice's private input, $i$,
at the end of the protocol
Bob can change the value of his own input $j$ simply
by applying a unitary transformation
to his own quantum machine.\footnote{A similar
idea is used in the proof of
the impossibility of
bit commitment\cite{Mayers1,Mayers2,LC1,LC2}.
That Alice knows nothing about Bob's chosen bit automatically
implies that Bob can cheat successfully by applying a unitary
transformation to change the value of the bit even after
the completion of the commitment phase. Thus, the commitment is
fake.} Therefore,
the index $j$ in Bob's density matrix
$\rho^{i,j} $ is redundant
in the sense that different values of $j$ simply
correspond to representing the density
matrix $\rho^i$ in different bases.

With such a simplification, one can essentially argue that
$\rho^i$ is a simultaneous eigenstate of $f(i,j_1), f(i,j_2)
, \cdots, f(i,j_m)$ in the following manner:
With an input $j_1$, Bob can learn $f(i,j_1)$. This
implies that $\rho^i$ is an eigenstate of $f(i,j_1)$.
But Bob can cheat by changing the value of $j$ from
$j_1$ to $j_2$
in the last minute
to learn $f(i,j_2)$ instead. This means that
$\rho^i$ is also an eigenstate of $f(i,j_2)$.
By repeating this argument, one sees clearly that
$\rho^i$ is a simultaneous eigenstate of all the
measuring operators for $f(i,j_1), f(i,j_2), \cdots,
f(i,j_m)$. Consequently, Bob can learn the values of
$f(i,j)$ for all values of $j$ simultaneously.
This is why the cheating strategy that I describe in
Subsection~2B works. In
the next Section, I will
generalize this attack to non-ideal protocols.

\section{non-ideal protocols}
A general non-ideal protocol may violate the
security requirements (a) and (b) slightly.
In relaxing (b), one would expect
that the unitary transformations that Bob uses for
changing $j$ from $j_i$ to $j_{i+1}$ to be imperfect.
In relaxing (a), the density matrix
that Bob has at the end of the protocol will now be slightly different from
an eigenstate of the measurement operator
that he uses. (This is because
Bob will generally be unable to determine the value of
$f(i,j_1)$ unambiguously in non-ideal protocols.)
Nonetheless, so long as the deviation from idealness is small,
one would expect Bob to learn a substantial
amount of
information about $f(i,j_2)$ even after his honest determination
of $f(i,j_1)$. That Bob can learn something about both $f(i,j_1)$
and $f(i,j_2)$ is already a serious violation of the security
requirement (c) and there is no need for one to consider
the security for $f(i,j_3)$, etc.
In other words,
one would expect that, for essentially the same reason
as the ideal protocol,
even non-ideal quantum one-sided
two-party computations are impossible. In what follows,
I prove that this is indeed the case. Readers who
are uninterested in technical details may skip the following
and go directly to Subsection~A.

More concretely, let me
relax security requirement (b) to allow Alice to have
a small probability to distinguish between different $j$'s.
I mimic the proof of Eq.\ (\ref{rot}).
As before, consider a dishonest Alice who tries to learn about $j$
by preparing an illegal initial state
${ 1\over \sqrt{n}} \sum_i | i \rangle_D \otimes | i \rangle_A $
where $n$ is the cardinality of $i$.
She keeps the first register, $D$, for herself and
uses the second register, $A$, to execute the two party protocol
honestly from this point on. Unlike the ideal case,
Eq.\ (\ref{Alice}) is violated for non-ideal protocols.
i.e., $ \rho^{Alice}_{j_1} \not= \rho^{Alice}_{j_2}$.
Nonetheless, so long as the probability for Alice to distinguish
successfully between the
two cases remains small, the two density matrices
$ \rho^{Alice}_{j_1}$ and $\rho^{Alice}_{j_2}$ must in some
sense be close to each other.

Mathematically,
the closeness between two density matrices $\rho$ and $\rho'$
of a system $S$ can be
described by the {\it fidelity}\cite{Jozsa}.
(See also Ref. \cite{Fuchs}.) Imagine another system $E$ attached to
a given system $S$. There are many pure states
$| \psi \rangle $ and $| \psi' \rangle $ on the composite system
that satisfy
\begin{equation}
{\rm Tr}_E \left( | \psi \rangle \langle \psi | \right) = \rho
\mbox{~~~~~~and~~~~~}
{\rm Tr}_E \left( | \psi' \rangle \langle \psi' | \right) = \rho'.
\label{pure}
\end{equation}
The pure states $| \psi \rangle $ and $| \psi' \rangle $
are called the {\it purifications} of the density
matrices $\rho$ and $\rho'$. The fidelity
$F( \rho, \rho')$ can be defined as
\begin{equation}
F( \rho, \rho') = max | \langle \psi | \psi' \rangle |
\label{fid}
\end{equation}
where the maximization is over all possible purifications.
I remark that\footnote{I thank R. Jozsa for a discussion
about this point.} for any fixed purification $\psi$ of $\rho$,
there exists a maximally parallel purification $\psi'$ of
$\rho'$ that satisfies Eq.\ (\ref{fid}).
Notice that $0 \leq F \leq 1$ and $F=1$ if and only if $\rho = \rho'$.

Returning to the discussion on non-ideal protocols,
the condition that the two density matrices $ \rho^{Alice}_{j_1}$ and
$\rho^{Alice}_{j_2}$ be close to each other can be specified by
the mathematical statement that
the fidelity
$F(  \rho^{Alice}_{j_1}, \rho^{Alice}_{j_2} )$
is close to $1$. Say
\begin{equation}
F(  \rho^{Alice}_{j_1}, \rho^{Alice}_{j_2} )
> 1 - \delta
\label{changej1}
\end{equation}
where $\delta  \ll 1 $.\footnote{One might imagine a situation when
Alice has been informed by her spy that Bob's input is either $j_1$
or $j_2$. In this case, her task is to distinguish between
these two remaining possibilities. To prevent Alice from
succeeding, it is crucial that Eq.\ (\ref{changej1}) holds.}
It follows from the definition of
fidelity in Eq. (\ref{fid})
that there exists a unitary transformation $U^{j_1 , j_2}$
acting on $H_B$ alone\footnote{A similar idea was
used by Mayers\cite{Mayers1} in
the discussion of non-ideal bit commitment schemes.}
such that
\begin{equation}
\left| \langle {v_j}_2 | U^{j_1, j_2} |{v_j}_1 \rangle \right|
> 1 - \delta .
\label{overlap}
\end{equation}
Since
(from Eqs.\ (\ref{vij}), (\ref{vj1}) and
(\ref{vj2}) ) $ |{v_j} \rangle = {1 \over \sqrt{n}}
\sum_i | i \rangle \otimes | v_{i j}\rangle $,
\begin{equation}
\left| \langle {v_j}_2 | U^{j_1, j_2} |{v_j}_1 \rangle \right|
=
{1 \over n} \left| \sum_i  \langle v_{i {j_2}} |U^{j_1, j_2}
| v_{i {j_1}} \rangle \right| 
> 1 - \delta .
\label{overlap1}
\end{equation}

Now
\begin{equation}
{1 \over n}  \sum_i \left|  \langle v_{i {j_2}} |U^{j_1, j_2}
| v_{i {j_1}} \rangle \right|
\geq {1 \over n} \left| \sum_i  \langle v_{i {j_2}} |U^{j_1, j_2}
| v_{i {j_1}} \rangle \right| 
> 1 - \delta .
\label{newoverlap1}
\end{equation}
For a protocol to be one-sided, one requires $\delta \ll 1$.
Let me consider the two cases:
(A) $ n \delta \ll 1$
and
(B) $ \delta \ll 1 \preceq n \delta$
separately.

{\it Case (A)}: $ n \delta \ll 1$.

It is a common requirement in computer science that $n \delta  \ll 1$.
In this case, for each $i$,
\begin{equation}
\left|  \langle v_{i {j_2}} |U^{j_1, j_2}
| v_{i {j_1}} \rangle \right| 
> 1 - n \delta 
\label{overlap2'}
\end{equation}
is still close to $1$.
I now come to the relaxation of security requirement
(a). Bob still chooses a $j$ say
$j_1$ and performs a measurement on his quantum state in
order to learn the value of $f(i,j_1)$.
However, for a non-ideal protocol, Bob's measurement result
will not give him full information on $f(i,j_1)$.
Nonetheless, for a protocol that is only slightly non-ideal,
one may demand that, for each $i$, Bob's
ignorance about $f(i,j_i)$ after his measurement would be much less
than one bit. That is to say that Bob's measurement can
extract the value of $f(i,j_1)$ from the density matrix
$\rho^{i,j_1}$ with a probability close to $1$.
Therefore, $\rho^{i,j_1}$ can be made to be almost an eigenstate of
Bob's measurement and thus the disturbance caused by such
measurement is small.
Consequently, one must have
\begin{equation}
F \Biggl(  \rho^{i, j_1}, {\cal E} \left( { \rho^{i, j_1}} \right)
\Biggr)  > 1 - \epsilon
\label{measure1}
\end{equation}
where $\epsilon \ll 1$ and
$ \cal E$ is a linear operator (the so-called
super-operator\cite{Knill})
which represents the action of
the (imperfect) measurement of $f(i,j_1)$ by Bob.
Since fidelity is preserved by unitary transformations,
one finds that
\begin{equation}
F \Biggl( U^{j_1, j_2} \rho^{i, j_1} \left( U^{j_1, j_2} \right)^{-1},
U^{j_1, j_2}   {\cal E} \left( \rho^{i ,j_1} \right) \left( U^{j_1, j_2}
\right)^{-1} \Biggr) > 1 - \epsilon .
\label{measure2}
\end{equation}
From Eqs.\ (\ref{overlap2'}) and (\ref{measure2}), one
deduces\footnote{This follows from
the fact that the fidelity is closely related\cite{Jozsa} to
the Bures metric.} that
\begin{equation}
F \Biggl( U^{j_1, j_2} {\cal E} \left( \rho^{i, j_1} \right) \left(
U^{j_1, j_2} \right)^{-1}, ~\rho^{i, j_2} \Biggr) >
 1 - O( n \delta) - O( \epsilon) .
\label{combine}
\end{equation}

Now the high fidelity of Eq.~(\ref{combine}) implies that
Bob's cheating strategy---of determining $f(i,j_1)$ approximately first,
applying a rotation to his state to change $j$ from $j_1$ to $j_2$
and
then determining $f(i,j_2)$---will allow him to
defeat the security
requirement (c)
of the protocol by learning substantial information about $f(i,j_2)$.
Therefore, even non-ideal protocols are unsafe if
$ n \delta \ll 1$.

{\it Case (B):} $ \delta \ll 1 \preceq  n \delta$.

I now separate the discussion further into two cases:
`typical' and `atypical' functions.
A `typical' function $f(i,j)$ is defined to be such that,
even if the value of $f(i,j_2)$ is
determined inaccurately by Bob for a small fraction say $1/10$ of the
$i$'s, Bob can still gain a considerable
amount of information about the value $i$.
With such a definition, I now argue that,
for a typical function, the assumption
$\delta \ll 1$ necessarily leads to a fatal violation of
security requirement (c), thus showing the insecurity of non-ideal protocols.
My point is the following:
Since each of the $n$ terms, $\left|  \langle v_{i {j_2}} |U^{j_1, j_2}
| v_{i {j_1}} \rangle \right|$, in Eq.\ (\ref{newoverlap1})
has a value less than or
equal to 1,
Eq.\ (\ref{newoverlap1})
implies that,
for at least nine out of ten of the $i$'s, the following is true:
\begin{equation}
\left|  \langle v_{i {j_2}} |U^{j_1, j_2}
| v_{i {j_1}} \rangle \right| 
> 1 - 10 \delta .
\label{overlap2}
\end{equation}
Since I am interested in Bob's density matrix,
I take the trace over Alice's quantum machine $A$ and
find that for each of those $i$'s,
\begin{equation}
F \left( U^{j_1, j_2} \rho^{i,j_1} \left( U^{j_1, j_2} \right)^{-1},
\rho^{ij_2} \right) 
> 1 - 10 \delta .
\label{overlap3}
\end{equation}

In relaxing the security requirement (a), Eqs.\ (\ref{measure1}) and
(\ref{measure2}) are still valid.
Combining Eqs.\ (\ref{measure2}) with (\ref{overlap3}),
one finds that for at least nine out of ten of the possible
$i$'s to be chosen by Alice,
\begin{equation}
F \Biggl( U^{j_1, j_2} {\cal E} \left( \rho^{i, j_1} \right) \left(
U^{j_1, j_2} \right)^{-1}, ~\rho^{i, j_2} \Biggr) >
 1 - O( 10 \delta) - O( \epsilon) .
\label{combine'}
\end{equation}
Hence, Bob can determine the value of $f(i,j_2)$ with
high accuracy
at least nine out of ten times. Since the function is
assumed to be typical, this implies that Bob can get substantial
amount of information about the value of $i$. Consequently,
the non-ideal protocol is insecure.

What about the case of `atypical' functions? An example of `atypical'
functions
is $f(i,j) = 1 $ if $i =j$ and $f(i,j) = 0$ otherwise (as in
quantum one-way oblivious identification in Corollary 2).
For those functions, it might be fatal if there exists a {\it single} $i$
such that
$\left|  \langle v_{i {j_2}} |U^{j_1, j_2}
| v_{i {j_1}} \rangle \right| $ is close to $0$.
In the above example, if $\left|  \langle v_{i {j_2}} |U^{j_1, j_2}
| v_{i {j_1}} \rangle \right| = 0 $ for $i =j_2$,
it might be the case that a cheating
Bob (after determining $f(i,j_1)$ honestly)
finds $f(i,j_2) = 0$ for all values of $i$.
Therefore, Bob gains no information about the value of
$i$ despite the high fidelity in Eq.\ (\ref{newoverlap1}).

I now argue that even atypical functions {\it cannot} be computed securely
in non-ideal one-sided secure computations whenever $\delta \ll 1$.
It is easiest to understand my reasoning by
working with an example. (I will present the general case
in two paragraphs below.)
Consider a situation in
which a cheating Alice prepares an unequally weighted
(i.e., non-maximally
entangled) state instead of an equally weighted (i.e., maximally entangled)
state in Eq.\ (\ref{Alicec}). For the function discussed
above ($f(i,j) =1$ if $i=j$ and $f(i,j)=0$
otherwise), suppose a cheating Alice prepares the state
$ { 1 \over \sqrt{2} } | j_2 \rangle_D \otimes
| j_2 \rangle_A +
{ 1 \over \sqrt{2 ( n -1)} } \sum_{ i \not= j_2} | i \rangle_D \otimes
| i \rangle_A  $ in her EPR attack (instead of Eq.\ (\ref{Alicec})).
Since Alice
is not supposed to learn much about Bob's input $j$, one must
still have $F( \rho^{Alice}_{j_1} , \rho^{Alice}_{j_2}) > 1 - \delta$.
This now implies that
\begin{equation}
\left|  \langle v_{j_2 {j_2}} |U^{j_1, j_2}
| v_{j_2 {j_1}} \rangle \right| > 1 - 2 \delta
\label{modi} ,
\end{equation}
and 
\begin{equation}
{ 1 \over n-1} \sum_{ i \not= j_2} 
\left|  \langle v_{i {j_2}} |U^{j_1, j_2}
| v_{i {j_1}} \rangle \right| 
> 1 - 2 \delta 
\label{modi1}
.
\end{equation}

Notice that the various $i$'s fall into two classes
(For $i=j_2$, $f(i,j_2) =1$. For $i \not= j_2$, $f(i,j_2) = 0$.)
which are to be distinguished by Bob.
Eq.\ (\ref{modi}) ensures that Bob will find
the value of $f(j_2, j_2)$ to be $1$ with high
probability. Similarly, Eq.\ (\ref{modi1}) ensures that Bob finds that
$f(i,j_2) $ to be zero with a high probability whenever $i \not= j_2$.
Therefore, Bob can determine with some confidence whether $i =j_2$
and it is clear that, for this particular example of $f(i,j)$,
even a non-ideal one-sided
secure computation is impossible.

Are secure one-sided computations impossible for {\it all} functions?
I now prove rigorously that they are impossible
for the case $\epsilon =0$ in Eq.\ (\ref{measure1}).
The
discussion for the case $\epsilon \not=0$ will be postponed to the very
end of this paragraph. 
When $\epsilon =0$, Bob determines the value of $f(i,j_1)$
accurately with certainty. Suppose he finds $f(i,j_1) $ to be $c$.
He can restrict his attention to the set, $S$, of $i$'s that
satisfy this constraint. If, for all pairs
$i,i'\in S$, $f(i,j) = f(i', j)$ for all $j$'s, then Bob
has nothing to gain in learning the value of $f(i,j_2)$.
Suppose the contrary. Then there exists a $j$ say $j_2$ such
that $f(i,j_2)$ is not a constant function in $S$.
Let me partition the set $S$ further into two or
more subsets $S_k$'s according to the value of $f(i,j_2)$.
Imagine a cheating Alice prepares a state
$ \sum_k { 1 \over |S_k|^{1/2} }
\sum_{i \in S_k} |i \rangle \otimes |i \rangle $.
Notice also that, with the above normalization
factor ${ 1 \over |S_k|^{1/2} }$,
each {\it set} $S_k$ in the partition is assigned equal weight
by Alice. (Here we ignore the obvious overall normalization factor.)
Such an assignment of weights maximizes the information gain
by Bob in performing his measurement. It is then
easy to see that so long as $\delta \ll 1$, Bob can
determine with some confidence to which
set $S_k$ $i $ belongs.
This seriously violates the security requirement (c).
In conclusion, I have shown rigorously
that secure one-sided computations are always
impossible for {\it any} function
when $\epsilon = 0$.
What about the general case when $\epsilon \not= 0$?
Since there
is no obvious singularity in the problem, provided
that $\epsilon $ is sufficiently
small, one-sided two-party secure computations should
remain impossible.

Notice that in the above
proof, I allow Bob's choice of
the unitary transformation to be dependent on the value
$f(i,j_1)$ that he has obtained. This is perfectly all right.

Finally, I remark that it is a matter
of definition that a {\it one-sided} protocol must have
$ \delta \ll 1$ in Eq.\ (\ref{changej1}). This is because
a protocol with $ \delta$ of order $1$ in Eq.\ (\ref{changej1})
is two-sided rather than one-sided.
For discussions on two-sided protocols, see next Section.

\subsection{Corollaries}
{\it Definition}: {\it One-out-of-two oblivious transfer} is an example of
one-sided two party secure computation in which the sender sends
two messages and the receiver chooses to receive
either message but cannot read both. Besides,
the sender, Alice, should not learn which message is read by the receiver,
Bob. More precisely,
Alice's input, $i$,
is a pair of messages, $(m_0, m_1)$ and Bob's input, $j$, is
a bit $0$ or $1$. At the end of the protocol, Bob learns
about the message $m_j$, but not the other message $m_{\bar j}$.
i.e., $f(m_0 ,m_1,j= 0) = m_0$ and $f(m_0 ,m_1,j= 1) = m_1$.

{\it Corollary~1}: Quantum one-out-of-two oblivious transfer is
impossible.

{\it Remark}: As noted in the introduction,
one-out-of-two oblivious transfer is
an important primitive for building up secure
computations. The impossibility of one-out-of-two oblivious transfer
itself is a major setback to quantum cryptography.
Also, this corollary is a generalization of Wiesner's
insight\cite{Wiesn} which showed that it is impossible
to achieve {\it ideal} quantum one-out-of-two
oblivious transfer using only {\it one-way} communications.

Incidentally, there have been claims that quantum cryptography
is useful for {\it one-way} oblivious identification\cite{CS,Huttner}.
Such a protocol would allow the first user Alice
to identify herself in front of a second user, Bob, by means of
a password, known only to both.
The safety requirement is that somebody, ignorant
of the password, impersonating Bob
shall not be able to
obtain much information on the password from the identification process.
One-way oblivious identification is
an example of one-sided two-party secure computation
in which the prescribed function
$f(i,j) = 1$ if $i=j$ and $f(i,j)=0$ otherwise.
In other words,
$f(i,j)$ gives a yes/no answer to the question whether the
two persons have the same password.
Such oblivious identification scheme is, therefore,
very useful for preventing frauds from typing PIN (Personal
Identificaton Number) to a dishonest teller machine that
steals passwords. 

{\it Corollary~2}: Quantum one-way oblivious identification is
impossible.

{\it Remark}: This result applies only to {\it one-sided}
schemes for quantum oblivious identification, a subject that
earlier papers\cite{CS,Huttner} have focused on and wrongly
claimed to achieve.
However, one should note that in practical applications,
assumption (b) in Section 2 can be relaxed.
For example, it is conceivable that one can allow the customer,
Alice, to learn substantial information about the input of Bob (the
cash machine). When Bob finds out in the computation that
someone is disguising herself as Alice
(the answer is `no' in the computation), he can
cancel Alice's password and ask Alice
to go to the bank in person to get a new password.
Such a protocol is much less powerful than what
the original protocols intend to achieve, but it is
still somewhat useful.
Also notice that the
possibility of {\it two-sided} schemes for oblivious identification
remains open. However, the following Section shows that
there exists a class of functions that cannot be computed
securely in any two-sided two-party secure computation.

\section{security of two-sided two-party computations}

{\it Definition}: Suppose Alice has a private input $i$
and Bob a private input $j$. A {\it two-sided two-party secure computation}
of a prescribed function $f(i,j)$ is a protocol
such that at the end,

(a) both Alice and Bob learn $f(i,j)$,

(b) Alice learns nothing about $j$ more than what logically
follows from $f(i,j)$ and her private input $i$, and

(c) Bob learns nothing about $i$ more than what logically
follows from $f(i,j)$ and his private input $j$.

Notice that in classical cryptography,
a one-sided two-party computation
of a function $f(i,j)$ can be reduced
to a two-sided two-party computation of
a function $F(i,j,r) = f(i,j)~~XOR~~r$ where $r$ is a random string of
input chosen by Bob and the XOR is taken bitwise.\footnote{I
thank R. Cleve for enlightening
discussions about this point.}
At the end of the protocol, both Alice and Bob learn
$F(i,j,r)$. While Bob can invert the function to
find  $f(i,j)=  F(i,j,r)~~XOR~~r$, Alice, being ignorant of
Bob's input $r$, has absolutely
no information about
$f(i,j)$.

Here I demonstrate explicitly
that the quantum two-sided two-party computation of
$F(i,j,r)$ is insecure.
Alice's density matrix at the end of
the protocol should only be a function of
$i$ and $F(i,j,r)$. This is because $F(i,j,r)$
is the only piece of information
that Alice is supposed to know about
Bob's inputs $j$ and $r$. Let me therefore
denote Alice's density matrix by
$\rho^{i,F(i,j,r)}_{Alice}$.
Suppose a dishonest Bob inputs $|j_1 \rangle \otimes
 { 1\over p^{1/2}}\sum_r |r \rangle \otimes  |r \rangle_D$ and
he keeps the system $D$ for himself.
(Here $p$ is the cardinality of $f(i,j)$, as $f(i,j) \in \{1,2, \cdots,
p \}$.)
In other words, he entangles the state of $r$ with a quantum dice $D$
and performs an EPR-type of cheating.
Suppose further that
a honest Alice inputs $i$. The density matrix that Alice has at the
end of the protocol
will simply be a normalized direct sum, $ { 1\over p} \sum_r
\rho^{i,F(i,j_1,r)}_{Alice}$,
of the individual density matrices.
For {\it any} fixed but arbitrary $j$, as $r$ changes,
$F(i,j,r)$ runs over all the $p$ values, $\{1, 2, \cdots , p \}$.
(Recall that $F(i,j,r) = f(i,j)~~XOR~~r$.)
Consequently, $ { 1\over p} \sum_r
\rho^{i,F(i,j_1,r)}_{Alice} = { 1\over p} \sum_r
\rho^{i,F(i,j_2,r)}_{Alice}$. i.e., Alice's density
matrix is {\it independent} of the value of $j$.
But then by precisely the same attack as in the one-sided case---by
determining
the value of $f(i,j_1)$, changing $j$ from $j_1$ to $j_2$
by a unitary transformation, determining the value of
$f(i,j_2)$ and so on,
Bob can determine the value of $f(i,j)$ for all values of $j$.
This violates the security requirement (c) for the two-sided protocol.
In conclusion, there are functions, namely $F(i,j_1,r)= 
f(i,j)~~XOR~~r$, that cannot
be computed securely by any two-sided protocol.

\section{Summaries and Discussions}
This paper deals with the applications of
quantum cryptography in the protection of
private information during public decision (rather than with the
most well-known application---so-called quantum key distribution).
As an important example, in a one-sided two-party secure computation,
one party Alice has a private input, $i$,
and the other party Bob
who has a private input, $j$. Alice helps Bob to
compute a prescribed function
$f(i,j)$ in such a way that at the end of the protocol,

(a) Bob learns $f(i,j)$,

(b) Alice learns nothing (or almost nothing) about $j$, 

and

(c) Bob knows nothing about $i$ more than
what logically follows from the value of $j$ and $f(i,j)$.

(For example, in password identification $f(i,j) = 1$ if $i=j$ and
$=0$ otherwise.)
Notice that Bob is supposed to choose
a $j$ (say $j_1$) and learn $f(i,j)$ for that particular
value of $j$ only. However, 
I prove that quantum one-sided two-party
computation is always insecure because Bob can learn
$f(i,j)$ for {\it all} values of $j$.
In the cheating strategy that I consider, Bob
determines the values of $f(i,j)$ for the various values of
$j$'s successively.\footnote{See footnote~8.}
That is to say that Bob inputs $j=j_1$, determines the
value of $f(i,j_1)$, changes $j$ to $j_2$ and
determines $f(i,j_2)$ and so on.

Such a cheating strategy works for
two reasons. For simplicity, let me first consider the
ideal protocol.
Let Bob input $j=j_1$
initially. Using the insight from
the impossibility of bit commitment\cite{Mayers1,Mayers2,LC1,LC2},
I prove that, owing to the security requirement (b),
Bob can cheat at the end of the protocol by changing the
value of $j$ from $j_1$ to $j_2$. Thus he can determine
the value of $f(i,j_2)$ instead of $f(i,j_1)$
as long as he has {\it not} performed a measurement to determine
$f(i,j_1)$ yet.
Of course, Bob is interested in learning $f(i,j_1)$ as well.
So, he must first measure the value of $f(i,j_1)$ before
rotating $j$ from $j_1$ to $j_2$.
If I can show that his measurement of $f(i,j_1)$ does not
disturb the quantum state he possesses, it is clear that
this cheating strategy will work. This is precisely what I do:
Since in an ideal protocol with an input $j=j_1$,
Bob can unambiguously determine the value of
$f(i,j_1)$ (security requirement (a)),
the density matrix that
Bob has must be an eigenstate of the measurement operator that
he uses. Consequently, he can measure the value of $f(i,j_1)$
{\it without} disturbing
the quantum state of the signal at all!
(Notice that, in effect, I have shown that owing
to the security requirements (a) and (b), the density
matrix that Bob has is a simultaneous eigenstate of
$f(i,j_1), f(i,j_2), \cdots, f(i,j_m)$. This contradicts
security requirement (c).)

These two points taken together mean that
this cheating strategy beats an ideal protocol
for one-sided two-party computation.\footnote{As discussed in the
introduction, one may also use the classical reduction theorem
from bit commitment to one-out-of-two oblivious transfer to argue
the impossibility of quantum one-sided two-party computations.
Such proof is, however, not transparent at all.
Yet another alternative
proof of the insecurity of {\it ideal} quantum one-sided two-party
computation can be made
by combining the idea of the proof of
the impossibility of quantum bit commitment
with a generalization of Wiesner's early
insight\cite{Wiesn} on the insecurity of a subclass of quantum
one-out-of-two oblivious transfer schemes.
Such proof is, however, non-constructive and does not apply directly to
non-ideal protocols. I shall, therefore, omit it here.}
In Section~4, I generalize my result to
show that a similar attack defeats non-ideal protocols
as well.
In conclusion, I have shown that
quantum one-sided two-party secure computation
(ideal or non-ideal) is always
impossible.

As corollaries to my results, contrary to popular
belief in earlier literature, quantum one-out-of-two
oblivious transfer and one-way oblivious
identification are also impossible.
I remark that the reduction theorem in classical cryptography
can be used to show that quantum (ordinary) oblivious transfer
is impossible. In future, it would be interesting to work out a direct
attack that defeats quantum oblivious transfer.

Since a one-sided two-party computation of a function
can be reduced to a two-sided two-party computation of
a related function,
there are functions that cannot be computed securely in two-sided
two-party computations
as well.
Can {\it any} function be computed securely in a quantum two-sided two-party
computation? While I do not have a definite answer,
the argument for
impossibility of ideal quantum coin
tossing\cite{LC2} can be used to prove the impossibility of {\it ideal}
two-sided
two-party secure computation (and also ideal so-called zero-knowledge proof).
Furthermore, Section~4 of Ref.~\cite{LC2} shows that
quantum two-sided two-party secure
computation can never be done
{\it efficiently}\footnote{Let me normalize everything so
that Alice and Bob both learn one bit of information from
executing a two-sided two-party computation.
If both parties are shameless enough
to stop running the protocol whenever one of them has an amount
of information that is $\epsilon$ greater than his/her opponent, it
is easy to show\cite{LC2} that the number $N$ of
rounds of communications needed for the protocol to be successful
has to satisfy $N \epsilon \geq  1$. An exponentially small $\epsilon$
requires an exponentially large $N$ and the scheme is necessarily inefficient.}
In conclusion, these results rule out the prefect
or nearly perfect protection of private information
in one-sided two-party computations by quantum mechanics.
The security of the quantum two-sided two-party computation
is also shown to be in very serious trouble.

In retrospect, there were good reasons for the reexamination of the
foundations of quantum cryptographic protocols such as
secure computation:
While
the security of quantum key distribution can
intuitively be attributed to the quantum no-cloning
theorem, no simple physical reason has ever been
given to the security of other quantum cryptographic
protocols such as bit commitment. This
is a highly unsatisfactory situation. Besides,
most proposed quantum
protocols are inefficient. From both theoretical and practical
points of view, a more fundamental understanding of
the issues of
security and efficiency of those protocols would therefore
be most welcome. 
In the claimed ``secure'' quantum bit commitment protocol\cite{BCJL},
researchers have implicitly assumed that measurements are made
by the two parties. What I have shown is that by using a quantum
computer and performing an EPR-type of attack,
the party, Bob, can defeat the security requirement of the
protocol. This is remarkable because the basic idea of the EPR attack
can be found in the pioneering papers\cite{Wiesn,BB84}.
The sky has fallen because its foundation has been shaky.

I emphasize that the cheating strategy proposed in this
paper generally requires a quantum computer to implement.
Before a quantum computer is ever built, quantum one-sided two-party
secure computations may still be secure in practice.
Besides, apart from quantum key
distribution (which is perfectly secure),
partial security provided by applications such
as quantum money may still be very useful.

On the positive side,
the impossibility of quantum one-sided two-party
computation together with the impossibility of
quantum bit commitment\cite{Mayers1,Mayers2,LC1,LC2}
constitute a major victory of cryptanalysis against {\it quantum}
cryptography.
On one hand, quantum key distribution is
secure because heuristically of the quantum no-cloning theorem.
On the other hand, quantum bit commitment
and quantum one-sided two-party
computation are impossible essentially because of the EPR paradox.
Therefore, there are now solid foundations to both
quantum cryptography and quantum cryptanalysis---the two sides
of the coin in quantum cryptology.
A key question remains as to the exact boundary to the power of
quantum cryptography. For instance, what is the power of quantum
cryptography in providing partial security in applications
such as quantum money?
Perhaps, new physical insights can
be gained in the attempts to answer this question.

\section{Acknowledgment}
I thank R. Cleve for encouraging me to tackle the
problem of secure computation and for enlightening discussions.
I am very grateful to
C. H. Bennett, G. Brassard,
C. Cr\'{e}peau, C. A. Fuchs,
L. Goldenberg, R. Jozsa, J. Kilian and
D. Mayers for numerous comments,
criticisms, discussions and references.
This work is a continuation of earlier work done together with H. F. Chau.
I thank him for fruitful collaborations.
I am also indebted to H. F. Chau and T. Spiller for their critical
readings of the manuscript.
Parts of the works were done during the
workshop on quantum computation held at
the Institute of Scientific
Interchange in Torino in June 96
and also during the quantum coherence and
quantum computation program at ITP, Santa Barbara in September 96.
Their hospitality
is gratefully acknowledged.
Finally, I thank S. Wiesner for his inspiring paper.
This research was supported in part by the National Science
Foundation under Grant No. PHY94-07194.

\end{document}